\documentclass[prd,twocolumn,superscriptaddress,altaffilletter,amssymb,showpacs,nofootinbib]{revtex4}
\usepackage[dvips]{graphicx}
\usepackage{amsmath}
\newcommand{\be}{\begin{equation}}
\newcommand{\ee}{\end{equation}}
\newcommand{\bea}{\begin{eqnarray}}
\newcommand{\eea}{\end{eqnarray}}

\begin{document}

\title{Thick DGP braneworlds}

\author{Israel Quiros}\email{israel@uclv.edu.cu}
\affiliation{Departamento de F\'{\i}sica, Universidad Central de
Las Villas, 54830 Santa Clara, Cuba.}\affiliation{Part of the Instituto
Avanzado de Cosmolog\'ia (IAC) collaboration http://www.iac.edu.mx/}

\author{Tonatiuh Matos}\email{tmatos@fis.cinvestav.mx}
\affiliation{Departamento de F{\'\i}sica, Centro de Investigaci\'on y de Estudios Avanzados del IPN,\\ A.P. 14-740, 07000 M\'exico D.F., M\'exico.} \affiliation{Part of the Instituto
Avanzado de Cosmolog\'ia (IAC) collaboration http://www.iac.edu.mx/}

\date{\today}
\begin{abstract}
We study Dvali-Gabadadze-Porrati braneworlds with finite thickness. In respect to standard (thin) DGP Friedmann equation, finite thickness of the brane causes a subtle modification of the cosmological equations that can lead to significant physical consequences. The resulting cosmology is governed by two length scales which are associated with the brane thickness and with the crossover length respectively. In this setup both early inflation and late-time acceleration of the expansion are a consequence of the 5D geometry. At early times, as well as at late times, 5D effects become dominant (gravity leaks into the extra dimension), while, at intermediate times, gravity is effectively 4D due to non-trivial physics occuring in standard (thin) DGP scenarios.
\end{abstract}

\maketitle

\section{Introduction}

Since the discovery that our universe can be currently undergoing a stage of accelerated expansion \cite{obs}, many phenomenological models based either on Einstein General Relativity (EGR), or using alternatives like the higher dimensional brane world theories
\cite{roy}, have been invoked (for a recent review on the subject see reference \cite{Copeland:2006wr}). The latter ones, being phenomenological in nature, are inspired by string theory.

One of the brane models that have received most attention in recent years is the so called Dvali-Gabadadze-Porrati (DGP) brane world \cite{dgp,dgpcosmology}. This model describes a brane with 4D world-volume, that is embedded into a flat 5D bulk, and allows for infrared (IR)/large scale modifications of gravitational laws. A distinctive ingredient of the model is the induced Einstein-Hilbert action on the brane, that is responsible for the recovery of 4D Einstein gravity at moderate scales, even if the mechanism of this recovery is rather non-trivial \cite{deffayet}. The acceleration of the expansion at
late times is explained here as a consequence of the leakage of gravity into the bulk at large (cosmological) scales, so it is just a 5D geometrical effect, unrelated to any kind of mysterious dark energy. 

Thin brane models are just an idealization of the physical reality. Braneworlds, if they are to be considered as models for our world, have to be of finite thickness.\footnotemark\footnotetext{However, there have been a few works on thick brane models in the bibliography. In \cite{mounaix}, for instance, a prescription for what to consider as a four-dimensional observable is given, and a Friedmann cosmology with time-indepenedent brane thickness is investigated. A similar scenario is studied in \cite{ghassemi} (see also \cite{navarro}).} The aim of the present paper is to show what the consequences are of considering finite-thickness DGP braneworlds within the cosmological context. Here we follow the prescriptions of reference \cite{mounaix}. It will be shown, in particular, that there arise both, ultraviolet (UV) and infra-red (IR) modifications of the laws of gravity. Actually, assumption of induced gravity on the finite-thickness brane leads to the existence of two different length scales, associated with the crossover length $r_c$ and with the brane thickness $\epsilon$ respectively. This fact hints at a possible (unified) geometrical description of early inflation and late-time accelerated expansion of the universe.

\section{The Model}

The thick brane model we are about to investigate rests upon the following assumptions (same assumptions as in reference \cite{mounaix}):

\begin{itemize}
\item{We consider a finite-thickness brane which is embedded in a five-dimensional (5D), Minkowski flat spacetime that is homogeneous and isotropic along three spatial dimensions.}
\item{Orbifold ($Z_2$) symmetry is assumed along the extra-dimension.}
\item{It is possible to find a Gaussian normal coordinate system centered on the middle layer of the brane.}
\item{Brane thickness is time-independent.}
\end{itemize}

The line element that respects the above listed assumptions is the following:

\bea &&ds^2_5=g_{AB}^{(5)}dx^Adx^B=-n^2(t,y)dt^2+\nonumber\\
&&\;\;\;\;\;\;\;\;\;\;\;\;\;a^2(t,y)\delta_{ij}dx^idx^j+dy^2,\label{line}\eea where, in our Gausian coordinate system, the extra-dimension is spanned by the coordinate $y$. The brane is localized between $y=-\epsilon/2$ and $y=\epsilon/2$ ($\epsilon$ stands for the time-independent brane thickness), and we use the following metric signature: $(-++++)$. 

The starting point will be the five-dimensional DGP action \cite{dgp,dgpcosmology}:

\bea S_{tot}=S_5^{EH}+S_4^{EH}+S_m,\label{action}\\
S_5^{EH}=-\frac{M_5^3}{2}\int d^5x\sqrt{|g_5|}\;R_5,\nonumber\\
S_4^{EH}=-\frac{M_4^2}{2\epsilon}\int d^4x\int dy\;\Delta_\epsilon(y)\;\sqrt{|g_4|}\;R_4,\nonumber\\
S_m=\int d^5x\sqrt{|g_5|}\;{\cal L}_m,\nonumber\eea where the delta function contribution $\delta(y)$ in the thin brane case has just been replaced by $\Delta_\epsilon/\epsilon$. The step-function $\Delta_\epsilon(y)$ is defined as it follows: $\Delta_\epsilon=1$, if $-\epsilon/2\leq y\leq\epsilon/2$, $\Delta_\epsilon=0$ otherwise (see \cite{griegos} for the same generalization of DGP models to thick brane contexts). It is worth noticing that, since in the limit $\epsilon\rightarrow 0$,

\be \lim_{\epsilon\rightarrow 0}\frac{\Delta_\epsilon(y)}{\epsilon}=\delta(y),\nonumber\ee then, in this (thin brane) limit, standard (thin) DGP model is recovered. 

In equation (\ref{action}), $S_5^{EH}$ is the five-dimensional Einstein-Hilbert action for the 5D metric $g_{AB}^{(5)}$, while $S_4^{EH}$ is the effective four-dimensional Einstein-Hilbert action. The 4D Ricci scalar $R_4$ is constructed from the four-dimensional metric induced on a given slice in the thick brane: $g_{AB}^{(4)}=g_{AB}^{(5)}-n_A n_B$ ($n_A$ is the normal to the slice). As customary $S_m$ is the action of the matter degrees of freedom. 

As it is stated in \cite{mounaix}, due to the non-null thickness of the brane, there is some arbitrariness in the definition of effective four-dimensional quantities that an observer living in the brane would measure. Here we follow the same prescription of \cite{mounaix} that is the simpler one can envisage. Given a 5D quantity $Q(t,y)$, we define its (spatial) average over the brane thickness $\bar Q(t)$ as follows:

\bea \bar Q(t)=\frac{1}{\epsilon}\int^{\epsilon/2}_{-\epsilon/2}dy\;\Delta_\epsilon(y)\; Q(t,y)=\nonumber\\
\frac{1}{\epsilon}\int^{\epsilon/2}_{-\epsilon/2}dy\; Q(t,y).\label{average}\eea It is the magnitude a 4D-observer living in the brane measures. In section \ref{4deff} we will discuss the implications of this prescription for the physical phenomena associated with the occurrence of two length scales in the model. 

The 5D Einstein's field equations that are derived from the action (\ref{action}) are the following ($k_5^2\equiv M_5^{-3}$):

\be G^A_{\;\;B}=k^2_5 T^A_{\;\;B}|_{Total}=k^2_5(T^A_{\;\;B}+U^A_{\;\;B}),\label{efe}\ee where, in correspondence with the symmetries assumed here, the stress-energy tensor for the matter degrees of freedom trapped to the thick brane is necessarily of the form\footnotemark\footnotetext{Since we have assumed the bulk spacetime to be Minkowski, then $T^A_{\;\;B}|_{Bulk}=0$.}

\be T^A_{\;\;B}=\frac{\Delta_\epsilon(y)}{\epsilon}\;\text{diag} [(-\rho_b,p_b,p_b,p_b),\epsilon P_T].\label{tbrane}\ee The other contribution to the energy-momentum $U_{AB}=(-\;M_4^2(\Delta_\epsilon/\epsilon)\;G_{\mu\nu}^4,\;0)$, comes from the four-dimensional scalar curvature $R_4$ induced in the thick brane. The non-null components of the latter contribution are:

\bea &&U^0_{\;\;0}=\frac{3M_4^2}{n^2 \epsilon}\;\Delta_\epsilon(y)\left(\frac{\dot a}{a}\right)^2,\nonumber\\
&&U^i_{\;j}=-\frac{M_4^2}{n^2 \epsilon}\;\Delta_\epsilon(y)\;\delta^i_j\left(-\left(\frac{\dot a}{a}\right)^2+2\frac{\dot a}{a}\frac{\dot n}{n}-2\frac{\ddot a}{a}\right).\label{utensor}\eea

\section{Cosmology with two length scales: Unifying Early-time and Late-time Inflation}

The basic equation determining the dynamics of the cosmic evolution can be written in the following form:\footnotemark\footnotetext{For details of the derivation of basic formulas see {\bf Appendix A}.}

\be \alpha^2 \epsilon^2\bar H^2=\left[-\;\kappa\;\eta+\frac{\epsilon^2}{2}\;(1+2\frac{r_c}{\epsilon})\bar H^2\right]^2+\frac{C \epsilon^2}{\bar a^4},\label{frw}\ee where $C$ is an arbitrary integration constant, and we have used the following definitions (see \cite{mounaix}):

\bea &&\bar H\equiv\frac{\bar{\dot a}}{\bar a},\;\;\;\kappa\equiv\frac{k_5^2}{6}\;\epsilon\bar\rho_b,\nonumber\\
&&\alpha\equiv\frac{a_{1/2}}{\bar a},\;\;\;\eta\equiv\frac{\bar{\rho_b a^2}}{\bar\rho_b \bar a^2}.\label{definitions}\eea The parameter $\kappa$ characterizes the thickness of the brane, while the quantities $\alpha$ and $\eta$ characterize the inhomogeneity of the brane along the fifth dimension. The crossover length is defined in the usual way:

\be r_c=\frac{M_4^2}{2M_5^3}.\nonumber\ee

If we compare equation (\ref{frw}) of the present section with equation (15) of reference \cite{mounaix} (assuming spatially flat cosmology with $k=0$, and a Minkowski bulk which means $\Lambda=0$), we notice that the only difference is in the factor $1+2r_c/\epsilon$ multiplying the averaged Hubble parameter $\bar H$ in the right-hand side (RHS) of (\ref{frw}). In what follows we shall explore the consequences of this tiny difference. 

From now on, until the contrary is specified, we assume the constant $C=0$, which amounts to ignoring the "dark" radiation term. This will make our analysis more transparent. Equation (\ref{frw}) can then be rewritten in the following way:

\be \bar H^2\mp\frac{2\alpha}{\epsilon+2r_c}\;\bar H=\frac{\bar k^2_4}{3}\;\bar\rho_b,\label{main}\ee where we defined the effective 4D gravitational constant measured by a brane observer

\be \bar k^2_4\equiv\frac{k^2_5\eta}{\epsilon+2r_c}.\label{newtonC}\ee  

Let us explore the limits of the modified Friedmann equation (\ref{main}). For $\epsilon\gg r_c$, we obtain:

\be \bar H^2\mp\frac{2\alpha}{\epsilon}\;\bar H=\frac{\bar k^2_4}{3}\;\bar\rho_b,\;\;\bar k^2_4=\frac{k^2_5\eta}{\epsilon},\label{modif1}\ee where the modified "crossover" length $\bar r_c\equiv \epsilon/2\alpha$, or

\be \bar r_c=\frac{\eta\; k^2_5/\bar k^2_4}{2\alpha}.\nonumber\ee For an infinite brane thickness ($\epsilon\rightarrow\infty$), supposing $\bar k^2_4$ is non-vanishing, we recover standard cosmology with

\be \bar H^2=\frac{\bar k^2_4}{3}\;\bar\rho_b.\nonumber\ee 

For $\epsilon\ll r_c$ the cosmic dynamics of the thick brane is dictated by the following modified Friedmann equation:

\be \bar H^2\mp\frac{\alpha}{r_c}\;\bar H=\frac{\bar k^2_4}{3}\;\bar\rho_b,\label{modif2}\ee where, now, the 4D gravitational constant measured by a brane observer is given by

\be \bar k^2_4=\frac{k^2_5\eta}{2 r_c}\label{g4}.\ee In respect to the thin DGP brane case, the effect of the brane thickness is to modify the crossover length ($\bar r_c\equiv r_c/\alpha$) and the strength of gravity (\ref{g4}), through the inhomogeneity of the brane along the extra dimension (quantified by the parameters $\alpha$ and $\eta$). In the thin brane limit $\epsilon\rightarrow 0$, since both $\alpha\rightarrow 1$ and $\eta\rightarrow 1$ (a thin brane is necessarily homogeneous in the extra-dimension), we recover standard DGP cosmology as it should be. Note that, in this case, $r_c=k^2_5/2\bar k^2_4$, so that, $\bar k_4^2=M_4^{-2}$. 

Let us focus on the so called "self-accelerating" branch of the Friedmann equation ("-" sign of the second term in the left-hand-side of equations (\ref{modif1}) and (\ref{modif2})). Assuming a vanishing averaged energy density in the thick brane $\bar\rho_b\rightarrow 0$, then, equations (\ref{modif1}) and (\ref{modif2}) lead to the following de Sitter expansion rates:

\be \bar H=\frac{2\alpha}{\epsilon},\;\;\;\bar H=\frac{\alpha}{r_c}.\label{asymptotes}\ee 

As seen, the two de Sitter expansion rates - as measured by a 4D observer - are driven by two different length scales: the brane thickness $\epsilon$, and the crossover length $r_c$, respectively.\footnotemark\footnotetext{Notice, from equations (\ref{asymptotes}), that the de Sitter expansion rate is also modified by the inhomogeneity of the thick brane along the extra dimension, being quantified by the parameter $\alpha$.} 

This feature of Thick DGP braneworld cosmology hints at the possibility that both, early time inflation, and late time accelerated expansion of the universe, could be explained as an effect of the leakage of gravity into the extra-space at early and late times respectively. However, for constant $r_c$ and $\epsilon$, this unification can not be accomplished in a consistent way. One possibility is to consider time-dependent brane thickness. A second possibility is to consider a running crossover length $r_c=r_c(t)$. The latter case could be associated with induced Brans-Dicke gravity on the thick brane, through replacement of $r_c\rightarrow r_0\;\phi$, where $\phi$ - the Brans-Dicke scalar field.

Within the context of the later possibility, a convenient scenario to address unified geometric description of early inflation and late time accelerated expansion could be the following. Assume the running crossover length $r_c(t)$ is a monotonically increasing function of the cosmic time ($0\leq r_c(t)\leq r_0$), such that, as expansion proceeds $r_c$ increases and, at late times, asymptotes to $r_0\gg \epsilon$. 

During the course of the early-time cosmic evolution ($r_c\ll \epsilon$), eventually, the stress-energy content dilutes and the universe settles in an early de Sitter stage: $\bar H=2\alpha/\epsilon$. This stage could be identified with the early inflationary period inherent in the standard cosmological model, so that, the brane thickness $\epsilon$ sets the scale at which early inflation happens. After the early stage of inflation, a mechanism for populating again the matter content of the universe is mandatory, however, here we do not aim at a discussion of this very delicate issue. 

Exit from inflation here is natural. Actually, as expansion further proceeds, the running crossover length further increases and, eventually becomes much larger than the brane thickness, so that, $\hat r=\epsilon+2r_c\rightarrow 2r_c\approx 2r_0$. As a consequence the cosmic dynamics is dictated now by the following Friedmann-DGP equation:

\be \bar H^2-\frac{\alpha}{r_0}\;\bar H=\frac{\bar k_4^2}{3}\;\bar\rho_b.\label{latetime}\ee For $1/r_0\ll\bar H\ll 1/\epsilon$ (recall that $r_0\gg \epsilon$), there is an intermediate stage after early inflation and before the onset of late-time accelerated expansion, where (approximate) standard Friedmann behaviour:

\be \bar H^2\approx\frac{\bar k_4^2}{3}\;\bar \rho_b,\ee drives the cosmic evolution. This stage corresponds to effective 4D (intermediate) regime, lasting between the two (early and late-time) inflationary stages. Eventually, as expansion further proceeds, the matter content dilutes with the expansion, so we are led with a late-time period of accelerated de Sitter expansion with $\bar H=\alpha/r_0$. From equation (\ref{latetime}) it is evident that, for late-time expansion to be accelerating, it is necessary that the crossover length $r_c\approx r_0>\alpha\bar H_0^{-1}$, where $\bar H_0$ is the present value of the Hubble parameter. Therefore, according to the present scenario of early inflation -- late-time accelerating expansion, it is required the brane thickness to be of the order of the inverse of the Hubble parameter during early inflation, while, the crossover length has to be, at least, of the order of the present value of the comoving Hubble radius ($\bar a_0=1$).

\begin{figure}[t!]
\begin{center}
\hspace{0.4cm}\includegraphics[width=6cm,height=4.5cm]{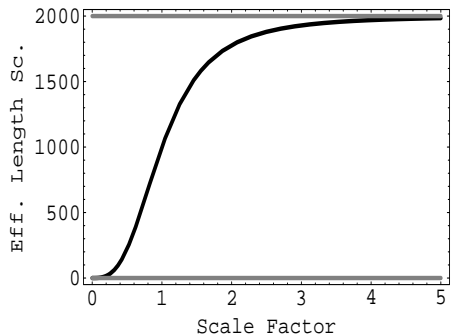}
\end{center}
\caption{Plot of the effective length scale $\hat r$ vs the averaged scale factor $\bar a$ for the toy example of section {\bf IV}. It is seen how, at early time (small $\bar a$-s), $\hat r\rightarrow \epsilon=0.1$ (lower gray line), while, at late times (large $\bar a$-s), $\hat r\rightarrow 2 r_c\approx 2 r_{0}=2000$ (upper gray line). The difference between the two length scales is roughly of four orders of magnitude, i. e., it has been significantly smoothed out in respect to the real physical situation.}
\label{fig1}
\end{figure}

\begin{figure}[t!]
\begin{center}
\hspace{0.4cm}\includegraphics[width=6cm,height=3.5cm]{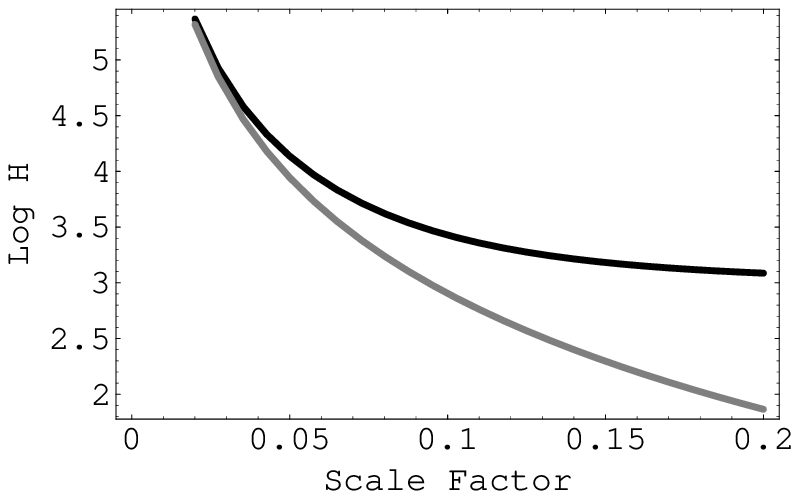}
\end{center}
\begin{center}
\hspace{0.4cm}\includegraphics[width=6cm,height=3.5cm]{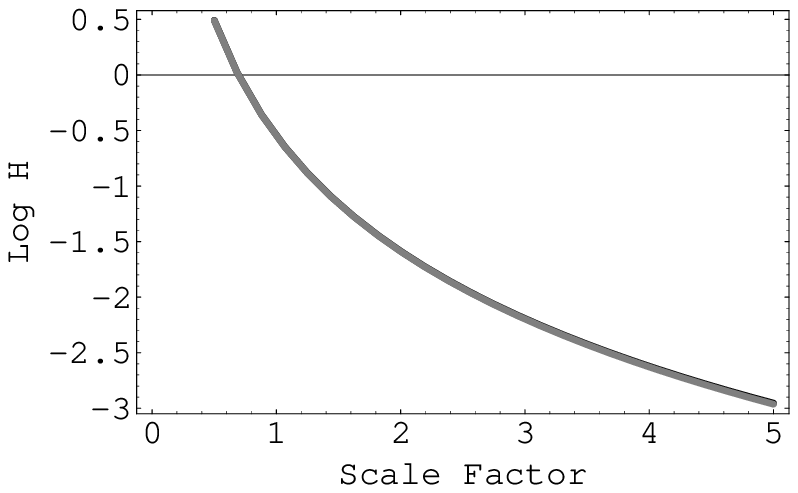}
\end{center}
\begin{center}
\hspace{0.4cm}\includegraphics[width=6cm,height=3.5cm]{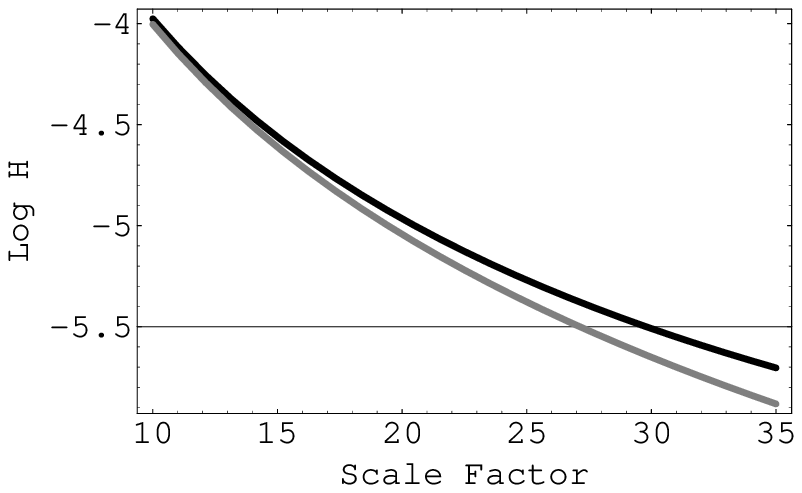}
\end{center}
\caption{Comparison between standard Friedmann behaviour (gray curve) and effective thick-DGP evolution of the (Log of the) averaged Hubble parameter (dark curve) for different stages of the cosmic expansion (early times - upper figure, intermediate times - figure in the center, late times - lower figure). Note that, before the onset of early inflation ($\bar a<0.05$), and at intermediate times ($0.5<\bar a<5$), the effective (averaged) cosmology on the thick-DGP brane is indistinguishable from standard 4D Friedmann behaviour.}
\label{fig2}
\end{figure}

\section{A Toy Example}

In order to illustrate the above discussion on the unified description of early inflation and late-time accelerated expansion as phenomena of purely geometric origin, here we shall study a toy example. We shall consider the following, ad hoc (arbitrarily chosen), evolution for the averaged brane matter energy density $\bar\rho_b$, and (running) crossover length $r_c$ (thick brane Brans-Dicke scalar field), respectively:\footnotemark\footnotetext{Here we consider the self-accelerating branch of the (thick) DGP braneworld only.}

\be \bar\rho_b=\rho_0\;\bar a(t)^{-3n},\;\;r_c=r_{0}\frac{\bar a(t)^m}{\bar a(t)^m+1},\ee where $n$, $m$, $\rho_0$, and $r_{0}$, are free parameters. The above arbitrary choice implies that the dynamics of the cosmic evolution is being fixed through the Einstein's field equations.

In the figure \ref{fig1} we show a plot of the effective length scale $\hat r=\epsilon+2r_c(t)$ vs the averaged scale factor $\bar a$. The free parameters have been (arbitrarily) chosen to be: $\rho_0=1$, $n=1$, $m=3$, $r_{0}=1000$, and $\epsilon=0.1$, respectively. Therefore, the difference between the two length scales roughly amounts to four orders of magnitude. In the real physical world this difference has to be much more significant since $\epsilon\sim\bar H_i^{-1}$ ($\bar H_i$ is the value of the averaged Hubble parameter at the end of inflation), while $r_0\sim\bar H_0^{-1}$ ($\bar H_0$ being the present value of the averaged Hubble parameter). However, for our purposes in this section, a difference of four orders of magnitude in the scales will be enough. 

Note that, at early times ($\bar a\rightarrow 0$), $r_c\rightarrow 0\;\Rightarrow \;\hat r\rightarrow \epsilon$, while, at late times (large $\bar a$), given $r_0\gg \epsilon$, $\hat r\rightarrow 2 r_c\approx 2 r_{0}$.  For $\bar a$-s in the range $0-0.4$, $\hat r\approx \epsilon$, so that early-time behaviour is dictated by the following Friedmann-DGP equation (a different writing of equation (\ref{modif1})):

\be \bar H_{e-time}=\frac{\alpha}{\epsilon}+\sqrt{\frac{\alpha^2}{\epsilon^2}+\frac{\bar k_4^2}{3}\;\bar\rho_b}.\label{early}\ee At late times ($\bar a$-s in the range $\bar a > 1$), on the contrary, the cosmic dynamics is driven by the modified Friedmann-Thick DGP equation (\ref{modif2}), that can be rewritten in the following way

\be \bar H_{l-time}=\frac{\alpha}{2r_c}+\sqrt{\frac{\alpha^2}{4r_c^2}+\frac{\bar k_4^2}{3}\;\bar\rho_b}.\label{het}\ee 

At intermediate times ($r_c\gg \epsilon$, $\bar\rho_b\gg \alpha^2/\bar k_4^2 r_c^2$, $\bar a$-s in the range $1<\bar a<10$), the cosmology is characterized by (approximate) standard Friedmann behaviour:

\be \bar H_{F}=\sqrt{\frac{\bar k_4^2}{3}\;\bar\rho_b}.\label{intermediate}\ee 

In the fig.\ref{fig2} we plot the (Log of the) thin DGP averaged Hubble parameter $\log\bar H_{e-time}$ (dark curve), and the (Log of the) standard one $\log\bar H_F$ (gray curve), vs the averaged scale factor $\bar a$, for the chosen values of the free parameters. Note that for $\bar a$-s in the range $0.05<\bar a<0.5$, the departure from standard Friedmann behaviour is apparent. For earlier times (prior to the onset of early inflation) there is a stage of standard (4D) Friedmann evolution that, notwithstanding, can be associated with standard Kaluza-Klein picture (infinite brane thickness limit). As expansion further proceeds ($1<\bar a<10$), the cosmic evolution is driven by a second (intermediate) period of effective 4D Friedmann evolution. Actually, note in the figure in the center of fig.\ref{fig2}, that thick-DGP expansion (dark curve), given by $\log\bar H_{l-time}$, and standard 4D Friedmann evolution (gray curve), given by $\log\bar H_F$, are indistinguishable. For $\bar a$-s larger than (approximately) $10$, the departure from standard Friedmann behaviour is again apparent. This illustrates in a toy example how the two (early-time and late-time) periods of inflation, are preceded by two periods of standard 4D Friedmann behaviour: one before the onset of early inflation, and a second one (the intermediate one) between the two inflationary stages. The origin of the different periods will be investigated in detail in the next section when we compute the corrections to the Newton's law of gravity arising from the occurrence of both length scales $r_c$ and $\epsilon$.

\section{Corrections to the Newton's Law}\label{4deff}

In standard (thin) DGP braneworlds with an infinite size extra dimension (Minkowski bulk), four-dimensional Newtonian gravity is recovered on small scales due to a very subtle mechanism. In fact, due to the infinite extent of the extra dimension, and to the fact that the bulk is Minkowski flat, there is no normalizable zero-mode graviton in the model. Four-dimensional gravity is reproduced as a resonance of the massive KK gravitons \cite{dgp,kazuya,koyama}. The massive graviton contains three additional degrees of freedom compared with a massless graviton. One of them is an extra scalar degree of freedom, so that, linearized gravity is described in this model by Brans-Dicke gravity with zero Brans-Dicke parameter. However, due to nonlinear shielding of the scalar mode, solar system constraints can be evaded \cite{koyama}.

In the present thick DGP brane model, in the thin brane limit $\epsilon\rightarrow 0$, or, equivalently, when $r_c\gg\epsilon$, four dimensional gravity is recovered in the way explained above. However, due to the possibility for $r_c$ to evolve in a cosmological set up, the opposite situation with $r_c\ll\epsilon$ has to be investigated as well. As we will inmediatelly show this possibility leads to new fenomena that could be associated with early time inflation. 

Since, in the present scenario, the laws of gravity are affected by two length scales: the crossover length $r_c=M_4^2/2M_5^3$ and the brane thickness $\epsilon$, we will pay due attention to the different limits where ultra-violet (UV) and infra-red (IR) modifications of these laws arise. In particular, corrections to the Newton's law of gravity become important in these limits.

The 5D Einstein's field equations that are derived from the action (\ref{action}) are the following (we omit here the matter-action term $S_m$):

\be G^5_{AB}=-\frac{2r_c}{\epsilon}\;\frac{\sqrt{|g_4|}}{\sqrt{|g_5|}}\;\Delta_\epsilon(y)G^4_{\mu\nu}\delta^\mu_A\delta^\nu_B.\label{efe1}\ee where, as already said, $r_c=M_4^2/2M_5^3$ is the crossover length. The only 4D Poincar\`e-invariant solution to (\ref{efe1}) is flat Minkowski space $\eta_{AB}=(\eta_{\mu\nu},1)$.

In order to understand how do corrections to the Newton's law arise in the thick DGP braneworld, it will be useful to explore both the ultra-violet (UV) and the infra-red (IR) limits of the graviton propagator computed from the action (\ref{action}), because, in these limits non-Newtonian corrections become important. As usual we simplify the analisys by ignoring the tensor structure, i. e., by considering only the scalar propagator $G(x,y)$. The classical equation for the Green's function looks like \cite{griegos}:

\be M_5^3\left(\nabla_5^2+\frac{2r_c}{\epsilon}\;\Delta_\epsilon(y)\;\nabla_4^2\right)G(x,y)=\delta^{(4)}(x)\delta(y),\ee where $\nabla_N^2$ is the flat Laplacian in $N$ dimensions. In Euclidean space we have:

\be M_5^3\left(p^2-\partial_y^2+\frac{2r_c}{\epsilon}\;\Delta_\epsilon(y)\;p^2\right)G(p,y)=\delta(y),\label{greeneq}\ee where $p\equiv\sqrt{p_4^2+p_1^2+p_2^2+p_3^2}$ ($p^0=-i p^4$).
The solutions to the latter equations are \cite{griegos}:

\be G(p,y)=A\;e^{-p|y|},\;\;\;|y|>\frac{\epsilon}{2},\ee outside of the thick DGP brane, while

\be G(p,y)=B\;e^{\hat p|y|}+C\;e^{-\hat p|y|},\;\;\;|y|\leq\frac{\epsilon}{2},\ee inside of the thickness of the brane. We have introduced the definition $\hat p\equiv\sqrt{1+2r_c/\epsilon}\;p$, while $A$, $B$, and $C$, are overall constants that can be found from Darmoix boundary conditions \cite{ghassemi}. In the present case these conditions amount just to continuity of the scalar propagator $G(p,y)$ and of its first $y$-derivative along the brane boundaries at $-\epsilon/2$ and $\epsilon/2$ respectively. We have:

\bea &&A=\frac{2\hat p}{p+\hat p}\;e^{(p-\hat p)\epsilon/2}\;C,\;\;\;\;B=\frac{\hat p-p}{\hat p+p}\;e^{-\hat p\epsilon}\;C,\nonumber\\
&&C=\frac{1}{2M_5^3\hat p}(1+\frac{p-\hat p}{p+\hat p}\;e^{-\hat p\epsilon})^{-1}.\label{constants}\eea Since the effective gravitational potential mediated by the scalar mode $G(p,y)$ is determined as:

\bea &&\bar V(r)=\int dt\; \bar G(t,r),\;\;\text{with}\nonumber\\
&&\bar G(x)=\int\frac{d^4p}{(2\pi)^4}\;e^{ipx}\;\frac{1}{\epsilon}\int_{-\epsilon/2}^{\epsilon/2}dy\;G(p,y),\label{effpot}\eea then the effective (averaged) propagator in Euclidean space will be given by the following expression:

\bea \bar G(p)=\frac{1}{\epsilon}\int_{-\epsilon/2}^{\epsilon/2}dy\;G(p,y)=\nonumber\\
\frac{2(C-B)}{\epsilon\;\hat p}+\frac{2}{\epsilon\;\hat p}\left(B\;e^{\hat p\epsilon/2}-C\;e^{-\hat p\epsilon/2}\right).\label{propa}\eea 

At this point we have to make a comment on the differences of our approach and the approach of reference \cite{griegos}. Actually, while in \cite{griegos} no a priori definition is given of what to consider as an effective 4D (observable) magnitude, in the present approach we follow the definition of \cite{mounaix} (basically equation (\ref{average})). In consequence we can define without ambiguities (in a way consistent with definition (\ref{average}) for averaged 4D quatities) the averaged (effective) 4D propagator in Euclidean space and, consequently, the effective 4D gravitational potential $\bar V(r)$ a four-dimensional observer trapped on the thick DGP brane would measure (see equation (\ref{effpot})).

Another comment has to do with what we understand as UV and IR limits in the present thick DGP model. Here we refer as the UV-limit to the case when the crossover length is much smaller than the brane thickness: $r_c\ll\epsilon$. This might be connected with the intuitive fact that, in this limit, the brane thickness $\epsilon$ has to play the role of the physically meaningfull scale at which 5D effects become important. Since, on the other hand, physical evidence tends to point to small enough $\epsilon$-s, then we can link this limit with a high-energy (short range) regime. On the oposite end it is the IR-limit (properly the standard thin DGP limit), i. e., the situation when the crossover length is much larger than the brane thickness: $r_c\gg\epsilon$. In this case arguments taken from cosmology point to large $r_c$-s ($r_c\sim H_0^{-1}$, where $H_0$ is the present value of the Hubble parameter). Therefore we can link the latter limit with a low-energy (long range) phase. 

We want to recall at this point that only a cosmological scenario where either $r_c$ or $\epsilon$ (or both) evolve in cosmic time, can bring us from the early UV-regime into the late-time IR-phase. In the present section we are considering constant $\epsilon$ and $r_c$. The cosmological arguments have been already considered in the former sections.

\subsection{UV-limit}

It is obvious that only in the limits when either $r_c\ll\epsilon$ or $r_c\gg\epsilon$, the corrections to Newton's law become important. In the intermediate regime the standard arguments on non-trivial physics mentioned at the beginning of this section apply, so that gravity is effectively four-dimensional, i. e., it is mediated by resonances of the KK gravitons. 

Let us first to consider the UV-limit, when $r_c\ll\epsilon$, i. e., the relevant scale is the brane thickness (this limit is missing in standard thin DGP models). In this case $\hat p\approx (1+r_c/\epsilon)\;p$, so that:

\bea &&B\approx\frac{r_c}{2\epsilon}\;e^{-p\epsilon}\;C,\nonumber\\
&&C\approx\frac{1}{2M_5^3 p}\left[1+\frac{r_c}{\epsilon}\left(1-\frac{e^{-p\epsilon}}{2}\right)\right]^{-1}.\label{uvconst}\eea At small momenta $p\ll 1/\epsilon$ (long wavelengths $\lambda\gg\epsilon$), from (\ref{propa}) it is seen that $\bar G(p)\approx B+C\sim C$, and, since, in this case,

\be C\approx\frac{1}{2M_5^3 p}\left(1-\frac{r_c}{2\epsilon}\right)\sim\frac{1}{2M_5^3 p},\nonumber\ee then, up to linear terms in $r_c/\epsilon$, the effective propagator $\bar G(p)\sim 2M_5^{-3} p^{-1}$ displays intrinsic 5D behaviour.

At large momenta $p\gg 1/\epsilon$ (short wavelengths $\lambda\ll\epsilon$), the overall constant $B\approx 0$ (see (\ref{uvconst})), while

\be C\approx\frac{1}{2M_5^3 p}\;(1-\frac{r_c}{\epsilon}),\nonumber\ee so that

\be \bar G(p)\approx\frac{2C}{\epsilon p}\;(1-\frac{r_c}{\epsilon})\approx\frac{1}{\epsilon M_5^3 p^2}\;(1-2\frac{r_c}{\epsilon}),\label{1+}\ee i. e., up to linear terms in $r_c/\epsilon$, $\bar G(p)$ is the Green's function for a four-dimensional theory.

\subsection{IR-limit}

This is properly the well-known thin DGP brane limit. In this case, since $r_c\gg\epsilon$, then $\hat p\approx\sqrt{2r_c/\epsilon}\;p$. It will be useful to introduce a couple of new variables: $\xi\equiv\sqrt{\epsilon/2r_c}\ll 1$, and $\sigma\equiv\sqrt{2\epsilon r_c}=\xi r_c$. It can be shown that, in terms of these variables, the effective propagator (\ref{propa}) can be written in the following form:

\be \bar G(p)=\frac{1}{\sigma M_5^3\;p^2}\;\left(1-\frac{\xi\;e^{-\sigma p/2}}{1-\left(1-\xi\right)\;e^{-\sigma p}}\right).\label{irpropa}\ee Two limiting cases are of importance:

\bigskip

(I) Small $\sigma p\ll 1$. 

\bigskip

This case corresponds to momenta $p\ll 1/\sqrt{2\epsilon r_c}$, i. e., wavelengths $\lambda\gg\sqrt{2\epsilon r_c}$. The propagator can be written in the following way:

\be \bar G(p)\approx\frac{1}{\sigma M_5^3 p^2}\frac{\sigma p}{\sigma p+\xi}.\label{ppg}\ee At very small momenta $p\ll\xi/\sigma=1/r_c$ (very large wavelengths $\lambda\gg r_c$), it approximates a 5D propagator

\be \bar G(p)\approx\frac{1}{\xi M_5^3 p}=\frac{\sqrt{r_c/2\epsilon}}{M_5^3 p}.\label{ppg1}\ee At intermediate momenta $\xi\ll\sigma p\ll 1\;\Rightarrow\;r_c^{-1}\ll p\ll 1/\sqrt{2\epsilon r_c}$ (intermediate wavelengths $\sqrt{2\epsilon r_c}\ll\lambda\ll r_c$), alternatively:

\be \bar G(p)\approx\frac{1}{\sigma M_5^3 p^2}=\frac{1}{\sqrt{2\epsilon r_c}M_5^3 p^2},\label{ppg2}\ee so that the propagator displays intrinsic four-dimensional behaviour.

\bigskip

(II) Very large $\sigma p\gg 1$. 

\bigskip

Here we are talking about very small wavelengths $\lambda\ll\sqrt{2\epsilon r_c}$. The effective propagator displays, again, 4D behaviour:

\be \bar G(p)\approx\frac{1}{\sigma M_5^3 p^2}.\ee Notice that this approximate expression for the propagator coincides with the one for the case of intermediate momenta (intermediate wavelengths $\sqrt{2\epsilon r_c}\ll\lambda\ll r_c$) in equation (\ref{ppg2}). Therefore, in general, for momenta $p\gg 1/r_c$ (wavelenghts $\lambda\ll r_c$) the propagator is given by the four-dimensional behaviour (\ref{ppg2}).

\section{Discussion}

As already pointed out, there is no way to have both UV and IR limits in the same theory, unless one considers cosmological evolution of either $r_c$ or $\epsilon$, or both. A possible cosmological scenario could be, for instance, to have a running $r_c$, that amounts to have Brans-Dicke (BD) gravity induced on the thick brane, with $r_c$ playing the role of the BD scalar field. 

Let us to summarise the main results of our investigation:

\begin{itemize}

\item In the UV regime gravity is effectively trapped on the thick DGP brane at length scales smaller that the brane thickness ($\lambda\ll\epsilon$). The effective 4D gravitational coupling is given by $\bar G_N=\bar M_4^{-2}$, where $\bar M_4^2=\epsilon M_5^3$. At length scales $\lambda$ much larger that $\epsilon$, gravity leaks into the extra space and 5D effects become important.

\item In the IR regime (properly thin DGP brane regime), due to resonances of the KK gravitons, gravity is effectively 4D at length scales $\lambda\ll r_c$, with effective gravitational coupling $\bar G_N=\sigma^{-1}M_5^{-3}$. At large (perhaps cosmological) scales $\lambda\gg r_c$, gravity leaks again into the extra space and 5D effects dominate.

\end{itemize}

Notice that there can be both a 4D as well as a 5D regimes associated with each limit, i. e., in a cosmological setup there could be four stages associated with passing from one limit into the other one: i) effective 4D behaviour with gravitational coupling $\bar G_N=(\epsilon M_5^3)^{-1}$ at very early times (UV-stage where $r_c\ll\epsilon$), then ii) 5D effects dominate at length scales $\lambda\gg\epsilon$. As $r_c$ further evolves with the cosmic expansion one goes from the UV-regime into the IR (standard, thin DGP brane) regime ($r_c\gg\epsilon$). Once the IR stage is reached, gravity is effectivelly 4D with $\bar G_N=\sigma^{-1}M_5^{-3}$ at length scales smaller than the crossover length $\lambda\ll r_c$ (this 4D stage is just a continuation of the intermediate one). Finally, at late times, the Universe enters a stage where 5D effects dominate at very large (perhaps cosmological) scales $\lambda\gg r_c$. 

We think this can be a nice cosmological scenario to address, in a unified frame, both early inflation and late-time accelerated expansion as phenomena of purely geometrical origin.

\section{Conclusions}

Consideration of finite thickness leads to a very convenient modification of the DGP scenario: there arise two length scales associated with the brane thickness and with the crossover length respectively. A central role in the present approach to thick DGP braneworlds is played by the prescription of what to consider as a four-dimensional observable \cite{mounaix}. According to this precription, spatial average in respect to the extra-dimension is defined through integration over the brane thickness, unlike standard dimensional reduction where integration is performed over the whole extent of the extra-dimension. In the present setup, as in standard thin DGP braneworlds, four-dimensional gravity is mediated by resonances of the KK gravitons. Massive KK modes lead to corrections of the Newton law that are appreciable as UV and IR limits are attained.

These corrections lead us to the conclusion that the short and large range behaviours of the laws of gravity depend on whether length scale dominates: $r_c$ or $\epsilon$. Accordingly, there can be two different Newtonian ($\bar V\sim 1/r$) regimes: one at $\lambda\ll\epsilon$ ($r_c\ll\epsilon$), and the other one at $\lambda\ll r_c$ ($r_c\gg\epsilon$). In the same way, one can find two different "five-dimensional" regimes: one at $\lambda\gg\epsilon$ ($r_c\ll\epsilon$), and the other one at $r\gg r_c$ ($r_c\gg\epsilon$). The existence of two stages of "five-dimensional" behavior is a very convenient fact to accomodate both early inflation, and late-time accelerated expansion of the cosmic evolution, within a unique geometrical picture where both inflationary stages are a consequence of the leakage of gravity into the extra-space.

\acknowledgements We gratefuly acknowledge useful comments by R. Maartens. This work was partly supported by CONACyT M\'exico, under grants 49865-F, 54576-F, 56159-F, and by grant number I0101/131/07 C-234/07, Instituto Avanzado de Cosmologia (IAC) collaboration. I. Q. acknowledges also the MES of Cuba for partial support of his research.

\appendix

\section{Deriving Friedmann's Equation}

In this Appendix we use definitions previously introduced in the main text (see for instance (\ref{definitions})). 

In terms of the metric (\ref{line}) the components of the five-dimensional Einstein's tensor $G_{AB}$ are:

\bea 
&&G_{00}=3\left\{\left(\frac{\dot a}{a}\right)^2-\frac{n^2}{\epsilon^2}\left(\left(\frac{a'}{a}\right)^2+\frac{a''}{a}\right)\right\},\nonumber\\
&&G_{ij}=\frac{a^2}{\epsilon^2}\;\delta_{ij}\left\{\frac{a'}{a}\left(\frac{a'}{a}+2\frac{n'}{n}\right)+2\frac{a''}{a}+\frac{n''}{n}\right\}+\nonumber\\
&&\;\;\;\;\;\;\;\;\;\;\;\;\;\;\;\;\;\;\;\;\;\;\frac{a^2}{n^2}\;\delta_{ij}\left\{\frac{\dot a}{a}\left(-\frac{\dot a}{a}+2\frac{\dot n}{n}\right)-2\frac{\ddot a}{a}\right\},\nonumber\\
&&G_{05}=3\left\{\frac{n'}{n}\frac{\dot a}{a}-\frac{\dot a'}{a}\right\},\nonumber\\
&&G_{55}=3\left\{\frac{a'}{a}\left(\frac{a'}{a}+\frac{n'}{n}\right)\right\}-\nonumber\\&&\;\;\;\;\;\;\;\;\;\;\;\;\;\;\;\;\;\;\;\;\;\;\;\;\;\;\;\;\;\;3\frac{\epsilon^2}{n^2}\left\{\frac{\dot a}{a}\left(\frac{\dot a}{a}-\frac{\dot n}{n}\right)+\frac{\ddot a}{a}\right\}.\eea Then, the $05$ component of Einstein's field equations (\ref{efe}) yields:

\be n(t,y)=\xi(t)\dot a(t,y).\ee We take the normalization $\bar n=1$, so that:

\be \xi=\bar{\dot a}^{-1}\;\;\Rightarrow\;\;n=\dot a/\bar{\dot a}.\ee On the other hand, since

\be T^0_{\;\;0}|_{Total}=\Delta_\epsilon\left(-\frac{\rho_b}{\epsilon}+\frac{3M_4^2}{\epsilon\; a^2}\;\left(\frac{\dot a}{n}\right)^2\right),\ee then, from $G^0_{\;\;0}=k_5^2 T^0_{\;\;0}|_{Total}$ it follows that:

\be \left(a^2\right)''=-\frac{2}{3}k_5^2 \epsilon\Delta_\epsilon\rho_b a^2+2 \epsilon^2\;\bar{\dot a}^2\left(1+2\frac{r_c}{\epsilon}\;\Delta_\epsilon\right).\ee where we have defined the crossover length $r_c=M_4^2/2M_5^3$. Integrating the last equation over the brane thickness one obtains:

\be a(1/2)'=\frac{\bar a}{\alpha}\left\{-\;\kappa\;\eta+\frac{\epsilon^2}{2}\left(1+2\frac{r_c}{\epsilon}\right)\bar H^2\right\}.\label{1}\ee The next step is to notice that the $55$ component of Einstein's field equations (\ref{efe}) can be written in the compact form:

\be \dot F=\frac{2}{3}k_5^2 \dot a a^3 P_T,\;\;\frac{F}{a^2}=\left(\frac{a'}{a}\right)^2-\bar{\dot a}^2,\ee where the following relationship holds:

\be G^5_{\;\;5}=\frac{3}{2}\frac{\dot F}{\dot a a^3}.\ee If we impose the boundary condition $P_T(\pm 1/2)=0$, i. e., $\dot F(\pm 1/2)=0$, then, after time integration, one obtains:

\be \epsilon^2 \bar H^2=\left[\frac{a'(1/2)}{\bar a}\right]^2+\frac{C \epsilon^2}{\alpha^2\bar a^4},\ee where $C$ is an arbitrary integration constant. By substituting (\ref{1}) into the last equation, the following Friedmann equation can be derived:

\be \alpha^2 \epsilon^2 \bar H^2=\left[-\;\kappa\;\eta+\frac{\epsilon^2}{2}\left(1+2\frac{r_c}{\epsilon}\right)\bar H^2\right]^2+\frac{C \epsilon^2}{\bar a^4},\label{frw1}\ee or, since $\kappa=k_5^2 \bar\rho_b \epsilon/6$, then:

\be \alpha^2 \bar H^2=\left[-\frac{k_5^2}{6}\;\eta\;\bar\rho_b+\frac{\hat r}{2}\bar H^2\right]^2+\frac{C}{\bar a^4},\nonumber\ee where $\hat r\equiv \epsilon+2 r_c$ is an effective length scale. The later equation can be recast into the form of Eq. (\ref{main}) of the main text.

\section{The Limits}

As seen from (\ref{main}), the thin brane limit $\epsilon\rightarrow 0$ ($\hat r=2r_c$), $\alpha\rightarrow 1$, $\eta\rightarrow 1$, yields the celebrated Friedmann equation for DGP braneworlds:

\be \bar H^2\mp\frac{1}{r_c}\;\bar H=\frac{\bar k_4^2}{3}\;\bar\rho_b.\ee 

The so called Kaluza-Klein limit $\epsilon\rightarrow\infty$ ($\hat r\rightarrow\infty$), instead, yields to the following standard Friedmann behavior:

\be \bar H^2=\frac{8\pi\hat G}{3}\;\bar\rho_b,\ee where we have considered $8\pi\hat G\equiv k_5^2\eta/2\hat r$ to be non-vanishing in this limit.


\end{document}